\documentclass[aps,twocolumn,amsmath,amssymb,superscriptaddress,prl]{revtex4-1}

\usepackage[dvipdfmx]{graphicx}
\usepackage{bm}
\usepackage{color}
\usepackage{epstopdf}

\begin{document}

\title{Spin-orbit torques originating from bulk and interface in Pt-based structures}

\author{Hiroki Hayashi}
\affiliation{Department of Applied Physics and Physico-Informatics, Keio University, Yokohama 223-8522, Japan}

\author{Akira Musha}
\affiliation{Department of Applied Physics and Physico-Informatics, Keio University, Yokohama 223-8522, Japan}

\author{Hiroto Sakimura}
\affiliation{Department of Applied Physics and Physico-Informatics, Keio University, Yokohama 223-8522, Japan}
\affiliation{School of Materials and Chemical Technology, Tokyo Institute of Technology, Tokyo 152-8522, Japan}

\author{Kazuya Ando}
\email{ando@appi.keio.ac.jp}
\affiliation{Department of Applied Physics and Physico-Informatics, Keio University, Yokohama 223-8522, Japan}
\affiliation{Keio Institute of Pure and Applied Science (KiPAS), Keio University, Yokohama 223-8522, Japan}
\affiliation{Center for Spintronics Research Network (CSRN), Keio University, Yokohama 223-8522, Japan}

\begin{abstract}
We investigated spin-orbit torques in prototypical Pt-based spintronic devices. We found that, in Pt/Ni and Pt/Fe bilayers, the damping-like torque efficiency depends on the thickness of the Pt layer. We also found that the damping-like torque efficiency is almost identical in the Pt/Ni and Pt/Fe bilayers despite the stronger spin memory loss at the Pt/Fe interface. These results suggest that although the dominant source of the damping-like torque is the bulk spin Hall effect in the Pt layer, a sizable damping-like torque is generated by the interface in the Pt/Fe bilayer due to the stronger interfacial spin-orbit coupling. In contrast to the damping-like torque, whose magnitude and sign are almost identical in the Pt/Ni and Pt/Fe bilayers, the field-like torque strongly depends on the choice of the ferromagnetic layer. The sign of the field-like torque originating from the bulk spin Hall effect in the Pt layer is opposite between the Pt/Ni and Pt/Fe bilayers, which can be attributed to the opposite sign of the imaginary part of the spin-mixing conductance. These results demonstrate that the spin-orbit torques are quite sensitive to the electronic structure of the FM layer. 

\end{abstract}

\maketitle

\section{I. introduction}

Current-induced spin-orbit torques provide a promising strategy for the electrical manipulation of magnetization in metals, semiconductors, and insulators~\cite{RevModPhys.91.035004,miron2011perpendicular,liu2012spinScience,yu2014switching,kurebayashi2014antidamping,fukami2016magnetization,avci2017current,Kageyamaeaax4278,an2016spin,PhysRevLett.117.116602,an2018manipulation,qiu2015spin,Fan2016,yu2014switching,safranski2019spin,kondou2016fermi,kim2013layer}. The efficient manipulation of magnetization through the spin-orbit torques offers a path for ultralow power, fast nonvolatile magnetic memory and logic technologies~\cite{manipatruni2019scalable}. The spin-orbit torques arise from the transfer of orbital angular momentum from the lattice to the spin system, which results from spin-orbit coupling in the bulk and at the interface of heavy-metal/ferromagnet (HM/FM) bilayers~\cite{RevModPhys.91.035004}.

The bulk spin-orbit coupling in the HM causes carriers with opposite spins to scatter in opposite directions. The spin dependent scattering generates a spin current from a charge current, which is known as the spin Hall effect~\cite{dyakonov1971current,Hirsch:1999aa,Sinova,Murakami,kato2004observation,KimuraPRL,Wunderlich,AndoPRL,RevModPhys.87.1213,0034-4885-78-12-124501,hoffmann2013spin}. In the HM/FM bilayer, the angular momentum carried by the spin Hall current is transferred to the magnetization through the spin-transfer mechanism~\cite{RevModPhys.91.035004}. This angular momentum transfer induces a torque on the magnetization, which is expressed as $\bm{\tau}_{\rm{DL}}=\tau_{\rm{DL}} \mathbf{m}\times (\bm{\sigma}\times\mathbf{m})$, where ${\bf m}$ is the magnetization unit vector, $\bm{\sigma}$ is the unit vector along the spin polarization direction of the spin current, and $\tau_\mathrm{DL}$ is the magnitude of the torque. The torque of this form is referred to as a damping-like torque. The transfer of the angular momentum is generally imperfect at the interface partly because of a small component of the spins that rotate when they reflect from the interface~\cite{PhysRevB.87.174411}. The spin rotation at the interface gives rise to a torque in the form of $\bm{\tau}_{\rm{FL}}=\tau_{\rm{FL}} \bm{\sigma}\times\mathbf{m}$, which is referred to as a field-like torque.

The interfacial spin-orbit coupling in the HM/FM bilayer also generates the damping-like and field-like torques. At the interface with broken inversion symmetry, the spin-orbit coupling lifts the electron-spin degeneracy, and the spin angular momentum is locked on the linear momentum~\cite{rashba1960properties,manchon2015new}. Because of the spin-momentum locking, a charge current flowing parallel to the interface generates a nonzero spin accumulation~\cite{edelstein1990spin,bel2008magneto}. This process, called the Rashba-Edelstein or inverse spin galvanic effect, exerts a torque on the magnetization through the exchange coupling at the HM/FM interface~\cite{PhysRevB.78.212405,miron2010current,miron2011perpendicular}. Since the interfacial Rashba spin-orbit effective field induces the rotation of the spin accumulation, both field-like and damping-like torques can be generated by the current-induced spin accumulation and exchange coupling~\cite{kurebayashi2014antidamping,PhysRevB.91.134402,PhysRevB.92.014402,PhysRevB.90.174423,PhysRevLett.121.017202,PhysRevApplied.9.014022}. Although in this scenario, carriers are assumed to be confined to the two-dimensional interface, in the HM/FM bilayer, carriers are not confined but can be scattered across the interface. In this situation, the interfacial spin-orbit coupling can also generate an interfacial spin current that flows away from the FM/HM interface through interfacial spin-orbit scattering. The interfacial spin-orbit scattering generates both damping-like and field-like torques~\cite{PhysRevB.94.104420}.

Understanding the physics behind the generation of the spin-orbit torques is essential for the development of spin-orbitronic devices, as well as the fundamental understanding of spin-dependent transport in condensed matter. A wide range of experiments have demonstrated that the spin-orbit torques can be manipulated by materials and interface engineering in Pt-based structures~\cite{zhu2019variation,PhysRevB.91.214416,Aneaar2250,PhysRevApplied.8.024034,doi:10.1063/1.5084201,an2018giant}, where the spin-orbit torques are generally attributed to the strong spin-orbit coupling of Pt, the most widely studied source of spin currents. However, despite this progress, the origin of the spin-orbit torques is still unclear and controversial even in the prototypical spin-orbitronic device. A typical example is the field-like torque in Pt/Ni-Fe-alloy bilayers~\cite{Aneaar2250,PhysRevB.91.214416,PhysRevB.98.024402,fan2013observation}. The reported values vary significantly for the same system and even the sign, as well as the magnitude, is inconsistent in literature, implying an important role of the spin-orbit coupling and electronic structure of the FM layer in the generation of the spin-orbit torques.

In this paper, we show that the origin of the spin-orbit torques in the standard Pt/FM bilayer strongly depends on the choice of the FM. We found that the damping-like torque efficiency depends on the thickness of the Pt layer in Pt/Ni and Pt/Fe bilayers. 
We also found that the damping-like torque efficiency is almost identical in the Pt/Ni and Pt/Fe bilayers despite the stronger spin memory loss at the Pt/Fe interface. These results suggest that although the dominant source of the damping-like torque is the bulk spin Hall effect in the Pt layer, a sizable damping-like torque is generated by the interface in the Pt/Fe bilayer due to the stronger interfacial spin-orbit coupling. 
We also found that the sign of the field-like torque originating from the bulk spin Hall effect in the Pt layer is opposite between the Pt/Ni and Pt/Fe bilayers, which can be attributed to the opposite sign of the imaginary part of the spin-mixing conductance. Although the strong spin-orbit coupling of the Pt layer is expected to play an essential role in the prototypical Pt-based structure, these results demonstrate that the spin-orbit torques are quite sensitive to the electronic structure of the FM layer. These results provide a crucial piece of information for a fundamental understanding of the spin-orbit torques.

\section{II. experimental method}

\begin{figure}[tb]
\center\includegraphics[scale=1]{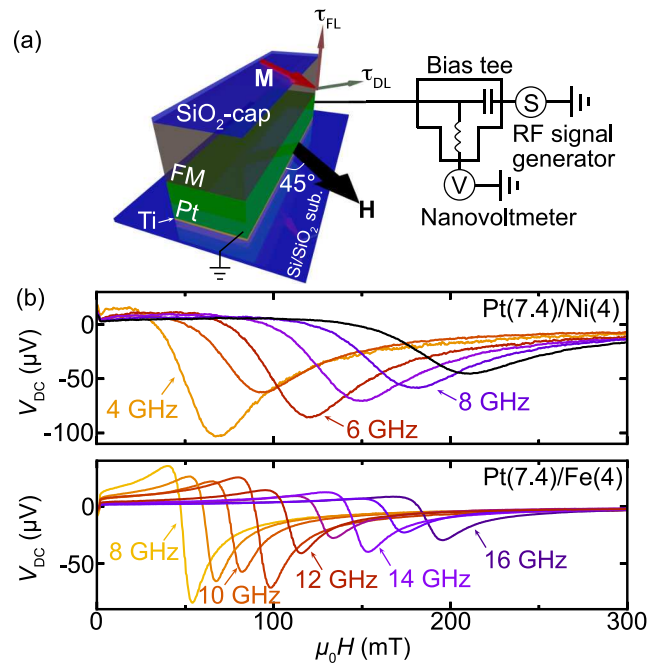}
\caption{
(a)~The schematic illustration of the experimental set up of the ST-FMR measurement for the Pt/FM ($\text{FM}=\text{Ni and Fe}$) bilayers. 
(b)~The magnetic field $H$ dependence of the DC voltage $V_{\text{DC}}$ for the Pt(7.4)/Ni(4) (upper) and Pt(7.4)/Fe(4) (lower) bilayers measured with the RF power of 24.7$\,\text{dBm}$. The RF frequency was varied from $f=4$ to $9\,\text{GHz}$ for the Pt/Ni film and from $f=8$ to $16\,\text{GHz}$ for the Pt/Fe film. The measurement of the ST-FMR for the Pt/Ni film at higher $f$ requires higher $H$, which limits the $f$ range in the present study. 
}
\label{fig1} 
\end{figure}

\begin{figure}[tb]
\center\includegraphics[scale=1]{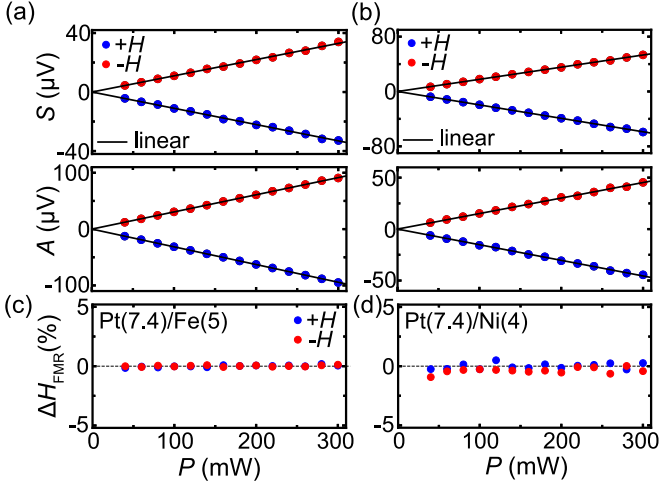}
\caption{The RF power $P$ dependence of the symmetric $S$ (upper) and antisymmetric $A$ (lower) components of the DC voltage $V_\mathrm{DC}$ for (a) the Pt(7.4)/Fe(5) bilayer at the RF frequency of $f = 10$~GHz and (b) the Pt(7.4)/Ni(4) bilayer at $f= 7$~GHz. The solid lines are the linear fitting result.  The RF power $P$ dependence of the change ratio of the FMR field $\Delta H_\mathrm{FMR} =  (H_\mathrm{FMR}(P) -  \bar{H}_\mathrm{FMR})/ \bar{H}_\mathrm{FMR}$ for (c) the Pt(7.4)/Ni(4) and (d) the Pt(7.4)Fe(5) bilayers, where $\bar{H}_\mathrm{FMR}$ represents the averaged value of $H_\mathrm{FMR}(P) $.}
\label{fig2} 
\end{figure}

\begin{figure}[tb]
\center\includegraphics[scale=1]{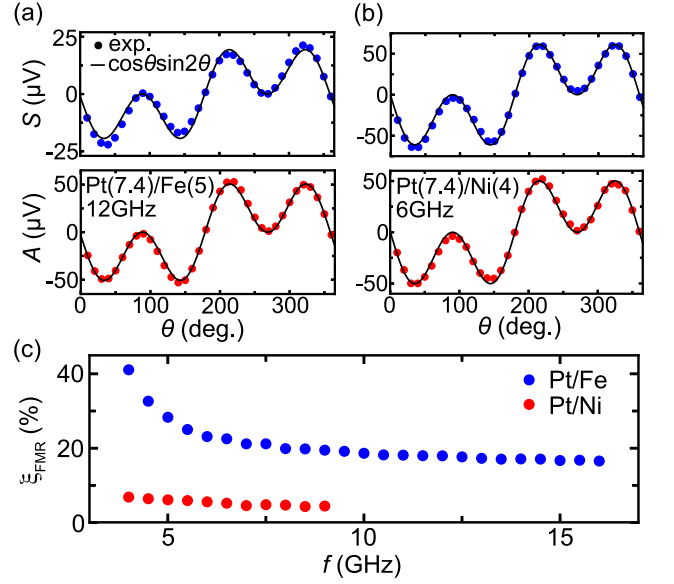}
\caption{The $S$ and $A$ components of the ST-FMR spectra as a function of the in-plane magnetic field angle $\theta$ for (a) the Pt(7.4)/Fe(5) film at $f$ = 12 GHz and (b) the Pt(7.4)Ni(4) film at $f$ = 6 GHz. Here, the in-plane external magnetic field $\mathbf{H}$ was applied at the angle of $\theta$ from the longitudinal direction. (c) The RF frequency $f$ dependence of the FMR spin torque generation efficiency $\xi_\mathrm{FMR}$ from $f =4$ to $16$ GHz for the Pt(7.4)/Fe(5) and from $f =4$ to $9$ GHz Pt(7.4)/Ni(5) bilayers.}
\label{fig3} 
\end{figure}

\begin{figure*}[tb]
	\center\includegraphics[scale=1]{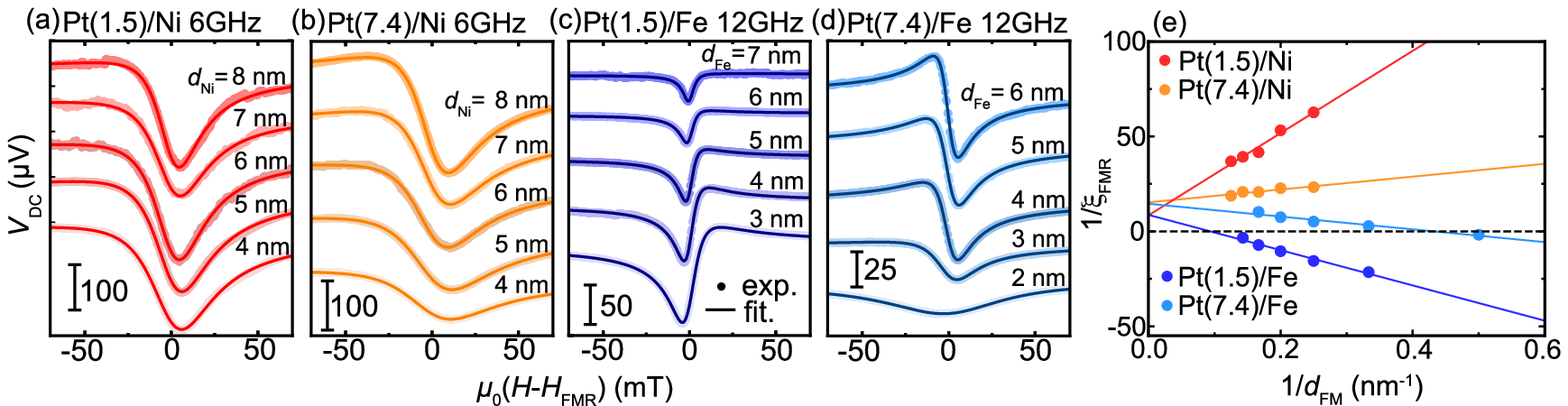}
	\caption{
	The magnetic field $H$ dependence of the DC voltage $V_{\text{DC}}$ for the (a)~Pt(1.5)/Ni($d_{\text{Ni}}$) at $f=6\,\text{GHz}$, (b)~Pt(7.4)/Ni($d_{\text{Ni}}$) at $f=6\,\text{GHz}$, (c) Pt~(1.5)/Fe($d_{\text{Fe}}$) at $f=12\,\text{GHz}$, and (d)~Pt(7.4)/Fe($d_{\text{Fe}}$) at $f=12\,\text{GHz}$. The solid circles are the experimental data and the solid curves are the fitting result using Eq.~(\ref{Lolentz}). (e)~The $1/d_{\text{FM}}$ dependence of $1/\xi_{\text{FMR}}$ for the Pt/Ni and Pt/Fe bilayers, where $d_{\text{FM}}$ is the thickness of the FM layer and $\xi_{\text{FMR}}$ is the FMR spin-torque efficiency at $f=9$~GHz for the Pt/Ni bilayer and $f=16$~GHz for the Pt/Fe bilayer. The solid circles are the experimental data and the solid lines are the fitting result using Eq.~(\ref{xi-thickness}).
	}
	\label{fig4} 
\end{figure*}

We investigated the spin-orbit torques in Pt/Ni and Pt/Fe bilayers using spin torque ferromagnetic resonance (ST-FMR). The sample structure is Ti(2)/Pt($d_{\text{Pt}}$)/FM($d_{\text{FM}}$)/SiO$_2$(5), where the numbers in the  parentheses represent the thickness in the unit of nm [see Fig.~\ref{fig1}(a)]. $d_{\rm{Pt}}$ and $d_{\rm{FM}}$ are the thickness of the Pt and FM layers, respectively. The films were deposited on thermally oxidized Si/SiO$_2$(100) substrates using RF-magnetron sputtering, where the base pressure was around $1\times 10^{-5}\,\text{Pa}$. The 2-nm-thick Ti layer was sputtered on the substrate, and then the Pt layer was sputtered on the adhesion Ti layer, where the deposition rate of Ti (Pt) was 0.01 (0.03) nm/s. On the Pt layer, the FM layer ($\text{FM}=\text{Ni and Fe}$) was sputtered, where the deposition rate of Ni (Fe) was 0.04 (0.02) nm/s. To avoid the natural oxidation of the FM layer, the 5-nm-thick SiO$_2$ was sputtered on the FM layer, where the deposition rate was 0.01 nm/s. All the sputtering processes were conducted in the 5N-purity argon atmosphere of 0.4 Pa at room temperature.

To measure the ST-FMR, the Pt/FM films were patterned into rectangular shapes (10 $\mu\rm{m} \times 150$  $\mu\rm{m}$) with Ti(2)/Pt(60) electrodes using the conventional photolithography and lift-off technique. For the ST-FMR measurement, we applied an RF current with the frequency $f$ along the longitudinal direction of the Pt/FM bilayer and an external magnetic field $\mathbf{H}$ at the angle of 45$^\circ$ from the longitudinal direction, as shown in Fig.~\ref{fig1}(a). The RF current generates out-of-plane and in-plane torques due to the spin-orbit torques and an Oersted field, which drive the magnetization precession under the FMR conduction. 
The precession of the magnetization leads to the oscillation of the resistance of the device through the anisotropic magnetoresistance (AMR) of the FM layer. In the bilayer, the mixing of the RF current and oscillating resistance induces DC voltage $V_{\rm{DC}}$. We measured $V_{\rm{DC}}$ for the Pt/FM bilayers using a bias tee and a nano-voltmeter at room temperature, as shown in Fig.~\ref{fig1}(a). Here, the phase of the RF current is almost constant in the ST-FMR devices because the wavelength of the RF current with the frequency of 4-16 GHz is much larger than the length of the devices, 150 $\mu$m. In fact, a previous study has demonstrated that the phase of the magnetization precession driven by an RF current is almost constant in a ST-FMR device with the width of 100 $\mu$m and the length of 400 $\mu$m~\cite{PhysRevApplied.11.034047}, supporting that the current non-uniformity is negligible in the present study.

The ST-FMR voltage $V_{\rm{DC}}$ can be decomposed into two components~\cite{PhysRevLett.106.036601}:
\begin{align}
V_{\text{DC}}=A\frac{W(\mu_0H-\mu_0H_{\text{FMR}})}{(\mu_0H-\mu_0H_{\text{FMR}})^2+W^2}\nonumber \\+S\frac{W^2}{(\mu_0H-\mu_0H_{\text{FMR}})^2+W^2},
\label{Lolentz}
\end{align}
where $W$ is the spectral width and $H_{\rm{FMR}}$ is the FMR field. The symmetric $S$ and anti-symmetric $A$ components are produced by the out-of-plane and in-plane effective fields, respectively. Here, the out-of-plane effective field is the damping-like effective field $H_\mathrm{DL}$ and the in-plane effective field is the sum of the Oersted field $H_\mathrm{Oe}$ due to the current flow in the Pt layer and the field-like effective field $H_\mathrm{FL}$. The damping-like and field-like torque efficiencies per applied electric field $E$, defined as 
\begin{equation}
\xi_{\text{DL(FL)}}^E=\left(\frac{2e}{\hbar}\right)\frac{\mu_0M_{\rm{s}}d_{\rm{FM}}H_{\text{DL(FL)}}}{E},
\end{equation}
can be determined by measuring the ST-FMR for devices with different $d_\mathrm{FM}$ using~\cite{PhysRevB.92.064426}
\begin{align}
	\frac{1}{\xi_{\mathrm{FMR}}}=\frac{1}{\xi_{\mathrm{DL}}^E}\left(\frac{1}{\rho_{\text{Pt}}}+\frac{\hbar}{e} \frac{\xi_{\mathrm{FL}}^E}{\mu_{0} M_{\mathrm{s}} d_{\mathrm{FM}} d_{\mathrm{Pt}}}\right),
	\label{xi-thickness}
\end{align}
where $e$ is the elementary charge, $\hbar$ is the Dirac constant, $\rho_{\rm{Pt}}$ is the electric resistivity of the Pt layer, and $M_s$ is the saturation magnetization. \begin{equation}
\xi_{\mathrm{FMR}}=\frac{S}{A} \frac{e \mu_{0} M_{s} d_{\rm{FM}} d_{\mathrm{Pt}}}{\hbar} \sqrt{1+ \frac{M_{\rm{eff}}}{ H_{\mathrm{FMR}}}}     \label{FMRefficiency}
\end{equation} 
is the FMR spin torque generation efficiency, where $M_{\rm{eff}}$ is the effective demagnetization field.
The resistivity $\rho_\mathrm{Pt} $ of the Pt layer in the Pt/FM bilayers was determined by the standard four-probe method for single Pt films, sputtered on the Ti layer, where the sheet resistance of the Ti layer was orders of magnitude larger than that of the Pt layer. The resistivity of the Pt layer is $\rho_\mathrm{Pt} = 65.6$ $\mu \Omega$cm for $d_\mathrm{Pt} = 1.5$ nm and $\rho_\mathrm{Pt} = 23.0$ $\mu \Omega$cm for $d_\mathrm{Pt} = 7.4$ nm. The increase of the resistivity induced by decreasing the thickness $d_\mathrm{Pt}$ can be attributed to the surface scattering, rather than a possible discontinuity of the Pt layer. In fact, the difference in the resistivity between the Pt layers with $d_\mathrm{Pt} = 1.5$ and 7.4~nm is in good agreement with that reported previously, where the drift-diffusion model is consistent with the experimental result for Pt/Co bilayers with $1.2\text{ nm} < d_\mathrm{Pt} < 15\text{ nm}$~\cite{PhysRevLett.116.126601}.

\begin{table*}[htb]
  \begin{center}
    \caption{The summarized parameters for the Pt/Ni and Pt/Fe bilayers, determined by the ST-FMR measurement. $\xi_{\rm{DL}}^E$ and $\xi_{\rm{FL}}^E$ are the damping-like and field-like torque efficiencies per applied electric field $E$, respectively. $K_\mathrm{s}$ and $\mu_0M_\mathrm{s}$ are the interfacial perpendicular magnetic anisotropy energy density and the saturation magnetization in the FM layer, respectively,  determined by fitting the FM-thickness dependence of the effective demagnetization field using Eq.~(\ref{K_s-fit}). $\text{Re} [G_{\text{eff,tot}}^{\uparrow\downarrow}]$ is the total effective spin mixing conductance, determined by fitting the FM-thickness dependence of the effective Gilbert magnetic damping $\alpha_{\rm{eff}}$ using Eq.~(\ref{G_eff-thick}). The errors are the standard deviation determined by the fitting.}
    \begin{tabular}{lcccccc} \hline\hline
       &$\xi_{\text{DL}}^{E}\,(\rm{10^3\Omega^{-1}cm^{-1}})$ & $\xi_{\text{FL}}^{E}\,(10^3\rm{\Omega^{-1}cm^{-1}})$ & $K_\mathrm{s}\,(\rm{mJ/m^2})$& $\mu_0M_\mathrm{s}\,\rm{(T)}$&$\text{Re} [G_{\text{eff,tot}}^{\uparrow\downarrow}]\,(10^{15}\rm{\Omega^{-1}m^{-2}})$\\ \hline 
      Pt(1.5)/Ni & $1.81\pm0.36$ & $0.42\pm0.02$& $0.05\pm0.02$  &$0.47\pm0.02$ & $1.30\pm0.25$\\
      Pt(7.4)/Ni & $2.84\pm0.12$ &  $0.53\pm0.05$& $0.14\pm0.01$ &$0.49\pm 0.01 $ &$1.73\pm0.08$\\
      Pt(1.5)/Fe & $1.75\pm0.15$ & $-0.77\pm0.03$& $1.41\pm0.04$  &$2.07\pm 0.01$ &$1.27\pm0.03$\\ 
      Pt(7.4)/Fe & $2.99\pm0.11$&  $-2.18\pm0.11$& $1.24\pm0.18$  &$1.93\pm 0.06$ &$1.26\pm0.05$\\ 
      \hline\hline
      \label{result}
    \end{tabular}
  \end{center}
\end{table*}

\section{III. results and discussion}

\subsection{A. Spin-torque ferromagnetic resonance}

In Fig.~\ref{fig1}(b), we show the magnetic field $H$ dependence of the DC voltage $V_{\rm{DC}}$ for the Pt(7.4)/Ni(4) and Pt(7.4)/Fe(4) bilayers. We first characterize the DC voltage $V_\mathrm{DC}$ for the Pt/FM devices by measuring RF current power $P$, magnetic field angle $\theta$, and RF current frequency $f$ dependence of the ST-FMR.

Figures~\ref{fig2}(a) and \ref{fig2}(b) show the RF power $P$ dependence of the ST-FMR signal for the Pt(7.4)/Fe(5) and Pt(7.4)/Ni(4) bilayers, where the symmetric $S$ and antisymmetric $A$ components of $V_\mathrm{DC}$ were obtained by fitting the measured $V_\mathrm{DC}$ using Eq.~(\ref{Lolentz}). This result shows that both symmetric $S$ and antisymmetric $A$ components are proportional to $P$, indicating that the ST-FMR is in the linear response regime. We also note that the FMR field $H_\mathrm{FMR}$ is independent of the applied microwave power $P$ as shown in Figs.~\ref{fig2}(b) and \ref{fig2}(c). The negligible change of $H_\mathrm{FMR}$ shows that the sample heating due to the microwave application and absorption barely affects the magnetic properties of the ST-FMR devices.

In Figs.~\ref{fig3}(a) and \ref{fig3}(b), we show the magnetic field angle $\theta$ dependence of the symmetric $S$ and antisymmetric $A$ voltage, where the in-plane external magnetic field $\mathbf{H}$ was applied at the angle of $\theta$ from the longitudinal direction of the Pt/Fe and Pt/Ni devices. In both films, $S$ and $A$ are proportional to $\cos\theta \sin 2\theta$, consistent with previous results ~\cite{fang2011spin}. This result indicates that the contribution from an out-of-plane Oersted field is negligible in the observed voltage, supporting the RF current flow is uniform in the ST-FMR devices.

The ST-FMR spectra at various RF current frequencies $f$ are shown in Fig.~\ref{fig1}(b). This result shows that the FMR field $H_{\rm{FMR}}$ changes systematically by changing $f$, which is consistent with the Kittel formula: $2\pi f=\gamma\sqrt{\mu_0H_{\rm{FMR}}(\mu_0H_{\rm{FMR}}+\mu_0M_{\rm{eff}})}$, where $\gamma$ is the gyromagnetic ratio. From the measured $V_\mathrm{DC}$, we calculated the FMR spin torque generation efficiency $\xi_\mathrm{FMR}$ at each $f$ using Eq.~(\ref{FMRefficiency}). The $f$ dependence of $\xi_\mathrm{FMR}$ for the Pt/FM bilayers is shown in Fig.~\ref{fig3}(c). Figure~\ref{fig3}(c) shows that $\xi_\mathrm{FMR}$ is almost independent of $f$ in the Pt/Ni bilayer, while $\xi_\mathrm{FMR}$ is independent of $f$ only at higher frequency in the Pt/Fe bilayer. Here, the ST-FMR predicts that $\xi_\mathrm{FMR}$ is independent of $f$~\cite{PhysRevLett.106.036601}. The result for the Pt/Ni bilayer is consistent with this prediction, showing that the measured $V_\mathrm{DC}$ is dominated by the ST-FMR for the Pt/Ni bilayer. The $f$ dependence of $\xi_\mathrm{FMR}$ for the Pt/Fe bilayer at the higher frequency is also consistent with the  prediction of the ST-FMR. In contrast, $\xi_\mathrm{FMR}$ at the lower frequency for the Pt/Fe bilayer cannot be entirely attributed to the ST-FMR. In the Pt/Fe bilayer, the increase of $\xi_\mathrm{FMR}$ at the lower frequency can be attributed to additional contributions, including thermal effects and the inverse spin Hall effect induced by spin pumping; because of the smaller AMR of the Pt/Fe film, compared to that of the Pt/Ni film, these additional effects can contribute to the observed $S$ in the Pt/Fe film. Since these contributions decreases with increasing $f$~\cite{doi:10.1063/1.3594661}, these additional contributions become negligible and the ST-FMR dominates the measured $\xi_\mathrm{FMR}$ at higher frequency $f$, resulting in the $f$-independent $\xi_\mathrm{FMR}$ in the Pt/Fe film. The negligible contribution from the spin pumping and thermal effects at the higher frequency is also supported by $1/d_\mathrm{FM}$ dependence of $1/\xi_\mathrm{FMR}$ as described below.  
In the following, we discuss the ST-FMR measured at the RF frequency $f$ where $\xi_\mathrm{FMR}$ is almost independent of $f$ to neglect the contribution from the spin pumping and thermal effects. 

In the following discussion, we neglect the spin-orbit torques arising from the FM/SiO$_2$ interface, since, in the Pt/FM/oxide structure, the contribution from the FM/SiO$_2$ interface to the spin-orbit torques is negligible compared to that from the Pt bulk and/or the Pt/FM interface~\cite{lee2019influences}.



\subsection{B. Spin-torque efficiencies}

In Figs.~\ref{fig4}(a) and \ref{fig4}(b), we show the $V_{\rm{DC}}$ spectra at $f= 6\,\rm{GHz}$ for the Pt(1.5)/Ni($d_\mathrm{FM}$) and Pt(7.4)/Ni($d_\mathrm{FM}$) bilayers with various $d_\mathrm{FM}$, respectively. We also show the $V_{\rm{DC}}$ spectra at $f= 12\,\rm{GHz}$ for the Pt(1.5)/Fe($d_\mathrm{FM}$) and Pt(7.4)/Fe($d_\mathrm{FM}$) bilayers in Figs.~\ref{fig4}(c) and \ref{fig4}(d), respectively. All the measured $V_{\rm{DC}}$ spectra are consistent with Eq.~(\ref{Lolentz}). We note that the sign of the anti-symmetric component $A$ is opposite between the Pt(1.5)/Fe and Pt(7.4)/Fe bilayers [see Figs.~\ref{fig4}(c) and \ref{fig4}(d)]. We also note that in the Pt(7.4)/Fe($d_\mathrm{FM}$) film, the sign of $A$ is reversed by decreasing $d_\mathrm{FM}$ from 3 nm to 2 nm. The sign changes are induced by the competition between the Oersted field $ H_{\rm{Oe}}$ and field-like effective field $H_{\rm{FL}}$, since $A\propto H_{\rm{Oe}}+H_{\rm{FL}}$.

To determine the damping-like and field-like torque efficiencies, $\xi_{\rm{DL}}^E$ and $\xi_{\rm{FL}}^E$, we plot $1/\xi_{\text{FMR}}$ at $f=9$~GHz for the Pt/Ni bilayer and $f=16$~GHz for the Pt/Fe bilayer as a function of $1/d_{\rm{FM}}$ in Fig.~\ref{fig4}(e). The measured values of $1/\xi_{\text{FMR}}$ is linear to $1/d_{\rm{FM}}$ in all the devices, consistent with Eq.~(\ref{xi-thickness}). 
The linear dependence of $1/\xi_{\text{FMR}}$ on $1/d_{\rm{FM}}$ supports that the additional contributions, including the spin pumping and thermal effects, are negligible in $\xi_{\text{FMR}}$. The reason for this is that in the presence of the additional contributions, $1/\xi_{\text{FMR}}$ deviates from the linear dependence on $1/d_{\rm{FM}}$~\cite{PhysRevApplied.14.024024}.

Figure~\ref{fig4}(e) shows that the sign of the intercept of the linear relation is positive in all the devices, showing that $\xi_{\rm{DL}}^E>0$ in the Pt/Ni and Pt/Fe bilayers because the intercept corresponds to $1/\xi_{\rm{DL}}^E$ [see Eq.~(\ref{xi-thickness})]. In contrast to the same sign of the intercept, the sign of the slope is opposite between the Pt/Ni and Pt/Fe bilayers. The slope of the linear relation corresponds to $\xi_{\rm{FL}}^E/\xi_{\rm{DL}}^E$, indicating that the sign of the field-like torque is opposite between the Pt/Ni and Pt/Fe bilayers: $\xi_{\rm{FL}}^E>0$ in the Pt/Ni film and $\xi_{\rm{FL}}^E<0$ in the Pt/Fe film. The values of $\xi_{\text{DL}}^E$ and $\xi_{\rm{FL}}^E$, obtained by fitting the data with Eq.~(\ref{xi-thickness}), are listed in Table~\ref{result}.

The result in Table~\ref{result} shows that the damping-like torque efficiency $\xi_{\rm{DL}}^E$ for the Pt($d_\mathrm{Pt}$)/Ni and Pt($d_\mathrm{Pt}$)/Fe bilayers is enhanced by increasing $d_\mathrm{Pt}$. This result also shows that $\xi_{\rm{DL}}^E$ is almost identical in the Pt($d_\mathrm{Pt}$)/Ni and Pt($d_\mathrm{Pt}$)/Fe bilayers. 
In contrast, the choice of the FM layer strongly affects the field-like torque; even the sign of $\xi_{\rm{FL}}^E$ is reversed by changing the FM layer from Ni to Fe. 

\subsection{C. Magnetic anisotropy and effective spin mixing conductance}
\begin{figure}[tb]
	\center\includegraphics[scale=1]{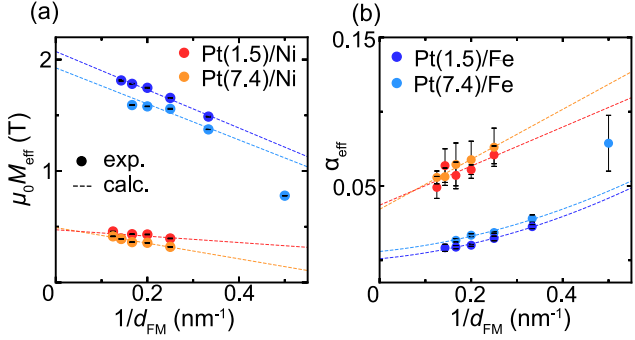}
	\caption{(a)~The $1/d_{\rm{FM}}$ dependence of the effective demagnetization field $\mu_0M_{\rm{eff}}$, where $d_\mathrm{FM}$ is the thickness of the FM layer. The dotted lines are the linear fitting result. (b) The effective magnetic damping $\alpha_{\rm{eff}}$ as a function of $1/d_{\rm{FM}}$. The dotted lines are the fitting result.
	}
	\label{fig5} 
\end{figure}

Before discussing the spin-torque efficiencies, we first characterize the spin-orbit coupling at the Pt/FM interface by quantifying the magnetic anisotropy and effective spin-mixing conductance, both affected by the interfacial spin-orbit coupling.
The interface magnetic anisotropy can be estimated from the $1/d_{\rm{FM}}$ dependence of $\mu_0M_{\rm{eff}}$ using~\cite{PhysRevB.92.214406}
\begin{align}
	\mu_0M_{\rm{eff}}=\mu_0M_{\text{s}}-\frac{2K_{\text{s}}}{M_{\text{s}}}\frac{1}{d_{\text{FM}}},
	\label{K_s-fit}
\end{align}
where $K_\mathrm{s}$ and $M_\mathrm{s}$ are the interface perpendicular magnetic anisotropy energy density and the saturation magnetization, respectively. 
We determined $\mu_0M_{\rm{eff}}$ by fitting the measured $f$ dependence of $H_\mathrm{FMR}$ using the Kittel formula.
In Fig.~\ref{fig5}(a), we show the $1/d_{\rm{FM}}$ dependence of $\mu_0M_{\rm{eff}}$. By fitting the data using Eq.~(\ref{K_s-fit}), we obtained $K_\mathrm{s}$ for the Pt/Ni and Pt/Fe films, as in Table~\ref{result}. The result shows that the magnitude of $K_\mathrm{s}$ at the Pt/Fe interface is an order of magnitude larger than that at the Pt/Ni interface, suggesting that the interfacial spin-orbit coupling at the Pt/Fe interface is stronger than that at the Pt/Ni interface~\cite{PhysRevLett.122.077201}.

The magnetic damping, affected by the interfacial spin-orbit coupling, also supports the stronger interfacial spin-orbit coupling at the Pt/Fe interface. In Fig.~\ref{fig5}(b), we show $1/d_\mathrm{FM}$ dependence of the effective Gilbert damping $\alpha_{\rm{eff}}$ for the Pt/Ni and Pt/Fe bilayers. The effective Gilbert damping $\alpha_{\rm{eff}}$ was determined by fitting the measured $f$ dependence of the ST-FMR linewidth $W$ using $W=W_0+(2\pi \alpha_{\rm{eff}}/\gamma)f$, where $W_0$ is the inhomogeneous linewidth. 
The effective Gilbert magnetic damping $\alpha_{\rm{eff}}$ in the presence of the spin pumping is expressed as~\cite{PhysRevLett.123.057203}
\begin{align}
\alpha_{\rm{eff}}=\alpha_{\text{int}}+\text{Re}[G_{\rm{eff,tot}}^{\uparrow\downarrow}]\frac{g\mu_{\text{B}} h}{4\pi e^2 M_{\text{s}}}\frac{1}{d_{\text{FM}}}+\frac{\beta_\mathrm{TMS}}{d_\mathrm{FM}^{2}},
\label{G_eff-thick}
\end{align} 
where $\alpha_{\rm{int}}$ is the intrinsic magnetic damping of the FM layer, $\text{Re}[G_{\text{eff,tot}}^{\uparrow\downarrow}]$ is the real part of a total effective spin mixing conductance, $g$ is the $g$-factor, $\mu_{\rm{B}}$ is the Bohr magnetron, and $h$ is the Planck constant. $\text{Re}[G_{\text{eff,tot}}^{\uparrow\downarrow}]$ consists of two components: $\text{Re}[G_{\text{eff,tot}}^{\uparrow\downarrow}]=\text{Re}[G_{\text{eff}}^{\uparrow\downarrow}]+G_{\text{SML}}$, where $G_{\text{eff}}^{\uparrow\downarrow}$ is the effective spin mixing conductance, which characterizes the spin relaxation in the bulk of the Pt layer, and $G_{\rm{SML}}$ characterizes the strength of the spin memory loss at the Pt/FM interface. 
In Eq.~(\ref{G_eff-thick}), the third term is the contribution from two-magnon scattering, where the coefficient $\beta_\mathrm{TMS}$ depends on both $(K_\mathrm{s}/M_\mathrm{s})^2$ and the density of magnetic defects at the FM surfaces~\cite{PhysRevB.60.7395,PhysRevB.62.5331,PhysRevLett.123.057203}. Since the two-magnon scattering at FM/oxide interfaces is known to be relatively weak, the primary source of the two-magnon scattering in the devices used in the present study is the Pt/FM interface. 
For the Pt/Fe bilayers, $\alpha_\mathrm{int}$ obtained by neglecting the $d_\mathrm{FM}^{-2}$ term is unphysically small or even negative, showing that the contribution from the two-magnon scattering cannot be neglected. In contrast, reasonable values of $\alpha_\mathrm{int}$ were obtained without taking into account the two-magnon scattering for the Pt/Ni bilayers. This difference is consistent with the fact that the strength of the two-magnon scattering depends on $(K_\mathrm{s}/M_\mathrm{s})^2$, which is clearly different in the Pt/Fe and Pt/Ni films, as shown in Table~\ref{result}. 
In Table~\ref{result}, we show $\text{Re}[G_{\rm{eff,tot}}^{\uparrow\downarrow}]$ for the Pt/FM bilayers, extracted by fitting the $1/d_{\rm{FM}}$ dependence of $\alpha_{\rm{eff}}$ in Fig.~\ref{fig5}(b) using Eq.~(\ref{G_eff-thick}). From the fitting, we obtained $\beta_\mathrm{TMS}=0.1$ nm$^2$ for the Pt/Fe bilayers, while we neglected the two-magnon scattering for the Pt/Ni bilayers because of the above reason. 

Table~\ref{result} shows that the effective spin-mixing conductance $\text{Re}[G^{\uparrow\downarrow}_{\rm{eff,tot}}]$ in the Pt/Ni bilayer depends on the Pt thickness $d_{\rm{Pt}}$, while $\text{Re}[G^{\uparrow\downarrow}_{\rm{eff,tot}}]$ in the Pt/Fe bilayer is independent of $d_{\rm{Pt}}$. 
Since the effective spin-mixing conductance $\text{Re}[G_{\text{eff}}^{\uparrow\downarrow}]$ depends on the thickness of the Pt layer $d_\mathrm{Pt}$, while $G_{\text{SML}}$ is independent of $d_\mathrm{Pt}$, 
this difference shows that the spin memory loss at the Pt/Fe interface is stronger than that at the Pt/Ni interface, supporting the stronger spin-orbit coupling at the Pt/Fe interface. This result is consistent with the stronger two-magnon scattering at the Pt/Fe interface because the two-magnon scattering increases with the strength of the interfacial spin-orbit coupling~\cite{PhysRevApplied.13.034038}. 

\subsection{D. Damping-like torque}
Next, we discuss the damping-like torque efficiency. The measured damping-like torque efficiency $\xi_{\rm{DL}}^E$ can be decomposed into $d_\mathrm{Pt}$-dependent $\xi_{\rm{DL,dep}}^E(d_{\rm{Pt}})$ and $d_\mathrm{Pt}$-independent $\xi_{\rm{DL,indep}}^E(d_{\rm{Pt}})$ components: $\xi_{\rm{DL}}^E(d_\mathrm{Pt})=\xi_{\rm{DL,dep}}^E(d_\mathrm{Pt})+\xi_{\rm{DL,indep}}^E$. 
One of the source of $\xi_{\rm{DL,dep}}^E(d_\mathrm{Pt})$ is the bulk spin Hall effect in the Pt layer. Although the damping-like torque due to the interfacial spin current originating from interfacial spin-orbit scattering also increases with $d_\mathrm{Pt}$~\cite{PhysRevB.94.104420,PhysRevApplied.7.014004}, 
we first neglect the contribution from this mechanism for simplicity.

The damping-like torque efficiency due to the bulk spin Hall effect in the simplest drift-diffusion model is expressed as $\xi_{\rm{DL,SHE}}^E (d_\mathrm{Pt})= [1-\mathrm{sech}(d_\mathrm{Pt}/\lambda_\mathrm{s}) ]\sigma_{\rm{SHE}}^\mathrm{eff}$, where $\sigma_{\rm{SHE}} ^\mathrm{eff}=\sigma_{\rm{SHE}} \varepsilon$ is the effective spin Hall conductivity of the Pt layer, where $\sigma_{\rm{SHE}} $ is the spin Hall conductivity and $\varepsilon$ represents the strength of the spin memory loss at the interface $( \varepsilon \le 1)$. 
The reason for using this simple model is that this study aims to clarify the role of the bulk and interface in the generation of the spin-orbit torques in the Pt/FM structures with the different FM layers, rather than to precisely quantify the bulk and interfacial spin-orbit torque efficiencies.

Since the spin diffusion length $\lambda_{\rm{s}}$ as well as the resistivity $\rho_{\rm{Pt}}$ of Pt films are dependent on the Pt thickness $d_\mathrm{Pt}$, we estimate $\lambda_{\rm{s}}$ for each Pt/FM devices with different $d_\mathrm{Pt}$, assuming the Elliot-Yafet spin relaxation mechanism, $\lambda_{\rm{s}}\propto 1/\rho_{\rm{Pt}}$: $\lambda_\mathrm{s}=0.98$~nm for $d_{\rm{Pt}}=1.5$~nm and $\lambda_\mathrm{s}=2.65$~nm for $d_{\rm{Pt}}=7.4$~nm, obtained from $\rho_{\rm{Pt}}\lambda_{\rm{s}}=0.61\times 10^{15}\,\rm{\Omega m^2}$~\cite{PhysRevB.94.060412}. For the Pt/FM films, we also assume that the spin Hall conductivity $\sigma_{\rm{SHE}} $ is independent of $d_\mathrm{Pt}$ because the Pt resistivity in the devices is in the moderately dirty regime~\cite{PhysRevB.94.060412}. Using the drift-diffusion model with the above assumptions, we obtain the change of the damping-like torque efficiency due to the spin Hall effect as $\xi_{\rm{DL,SHE}}^E(d_\mathrm{Pt}=7.4~\mathrm{nm})/\xi_{\rm{DL,SHE}}^E(d_\mathrm{Pt}=1.5~\mathrm{nm})=1.42$.  

From the values shown in Table~\ref{result}, we obtain $\xi_{\rm{DL}}^E(d_\mathrm{Pt}=7.4~\mathrm{nm})/\xi_{\rm{DL}}^E(d_\mathrm{Pt}=1.5~\mathrm{nm})=1.57\pm 0.13$ for the Pt/Ni bilayer and $\xi_{\rm{DL}}^E(d_\mathrm{Pt}=7.4~\mathrm{nm})/\xi_{\rm{DL}}^E(d_\mathrm{Pt}=1.5~\mathrm{nm})=1.71\pm 0.05$ for the Pt/Fe bilayer. 
The experimental value of $\xi_{\rm{DL}}^E(d_\mathrm{Pt}=7.4~\mathrm{nm})/\xi_{\rm{DL}}^E(d_\mathrm{Pt}=1.5~\mathrm{nm})$ for the Pt/Ni bilayer is close to the prediction of the above simple model of the spin Hall effect, suggesting that the damping-like torque in the Pt/Ni bilayer is dominated by the spin Hall effect. 
The experimental value of $\xi_{\rm{DL}}^E(d_\mathrm{Pt}=7.4~\mathrm{nm})/\xi_{\rm{DL}}^E(d_\mathrm{Pt}=1.5~\mathrm{nm})$ for the Pt/Fe bilayer is also comparable to the prediction of the spin Hall effect. However, it is reasonable to consider that a sizable damping-like torque is generated not only by the bulk spin Hall effect in the Pt layer but also by different mechanisms other than the spin Hall effect in the Pt/Fe bilayer. 
The reason for this is that the magnitude of $\xi_\mathrm{DL}^E$ of the Pt/Fe bilayer is comparable to that of the Pt/Ni bilayer despite the fact that the stronger spin memory loss at the Pt/Fe interface suppresses the damping-like torque due to the spin Hall effect. This suggests that the damping-like torque originating at the Pt/Fe interface contributes to $\xi_{\rm{DL}}^E$ in the Pt/Fe bilayer, which is consistent with that the interfacial spin-orbit coupling in the Pt/Fe bilayer is stronger than that in the Pt/Ni bilayer.

\subsection{E. Field-like torque}

The field-like torque efficiency $\xi_{\rm{FL}}^E$ can also be decomposed into $d_\mathrm{Pt}$-dependent $\xi_{\rm{FL,dep}}^E(d_\mathrm{Pt})$ and $d_\mathrm{Pt}$-independent $\xi_{\rm{FL,indep}}^E$ components: $\xi_{\rm{FL}}^E(d_\mathrm{Pt})=\xi_{\rm{FL,dep}}^E(d_\mathrm{Pt})+\xi_{\rm{FL,indep}}^E$. We note that $\xi_{\rm{FL,dep}}^E(d_\mathrm{Pt})$ and $\xi_{\rm{FL,indep}}^E$ mainly originate from the bulk spin Hall effect and interfacial spin-orbit coupling, respectively; the field-like toque originating from the spin current due to the interfacial spin-orbit scattering, as well as the spin accumulation due to the Rashba-Edelstein effect, is insensitive to $d_\mathrm{Pt}$~\cite{PhysRevB.94.104420,PhysRevApplied.7.014004}. The separation of the bulk and interface contributions from the result shown in Table~\ref{result} based on an established model requires to assume various parameters, such as the real and imaginary parts of the spin mixing conductance, as well as the spin Hall conductivity~\cite{PhysRevB.87.174411,PhysRevB.91.144412}, making the analysis less reliable. Thus, we use an alternative approach to roughly estimate the bulk and interface contributions to the field-like torque. We assume that the field-like torque efficiency due to the spin Hall effect, $\xi_{\rm{FL,SHE}}^E(d_\mathrm{Pt})$, follows the same $d_\mathrm{Pt}$ dependence as with the damping-like torque efficiency $\xi_{\rm{DL,SHE}}^E(d_\mathrm{Pt})$: $\xi_{\rm{FL,SHE}}^E(d_\mathrm{Pt})=[1-\mathrm{sech}(d_\mathrm{Pt}/\lambda_\mathrm{s})]\xi_{\rm{FL,SHE}}^{E,0}$ or $\xi_{\rm{FL,SHE}}^E(d_\mathrm{Pt}=7.4~\mathrm{nm})/\xi_{\rm{FL,SHE}}^E(d_\mathrm{Pt}=1.5~\mathrm{nm})=1.42$.

From the values shown in Table~\ref{result}, we obtain $\xi_{\rm{FL}}^E(d_\mathrm{Pt}=7.4~\mathrm{nm})/\xi_{\rm{FL}}^E(d_\mathrm{Pt}=1.5~\mathrm{nm})=1.26\pm 0.08$ for the Pt/Ni bilayer and $\xi_{\rm{FL}}^E(d_\mathrm{Pt}=7.4~\mathrm{nm})/\xi_{\rm{FL}}^E(d_\mathrm{Pt}=1.5~\mathrm{nm})=2.83\pm0.02$ for the Pt/Fe bilayer. This result shows that $\xi_{\rm{FL}}^E(d_\mathrm{Pt}=7.4~\mathrm{nm})/\xi_{\rm{FL}}^E(d_\mathrm{Pt}=1.5~\mathrm{nm})$ for the Pt/Fe film is clearly different from the expectation from the field-like torque efficiency $\xi_{\rm{FL,SHE}}^E(d_\mathrm{Pt})$ due to the bulk spin Hall effect. This shows that a sizable field-like torque is generated by the Pt/Fe interface, which is consistent with the strong spin-orbit coupling at the Pt/Fe interface.

In the Pt/Fe bilayer, the clear change of $\xi_{\rm{FL}}^E$ induced by changing $d_\mathrm{Pt}$ indicates that a sizable field-like torque is also generated by the bulk spin Hall effect. To test the role of  the bulk spin Hall effect in the field-like torque efficiency $ \xi_{\rm{FL}}^E$, we define the change of the field-like torque efficiency $\Delta \xi_{\rm{FL}}^E$ as 
$\Delta \xi_{\rm{FL}}^E= \xi_{\rm{FL}}^E(7.4~\mathrm{nm})-\xi_{\rm{FL}}^E(1.5~\mathrm{nm})=\xi_{\rm{FL,dep}}^E(7.4~\mathrm{nm})-\xi_{\rm{FL,dep}}^E(1.5~\mathrm{nm})$. From the result shown in Table~\ref{result}, we obtain $\Delta \xi_{\rm{FL}}^E>0$ for the Pt/Ni bilayer, while $\Delta \xi_{\rm{FL}}^E<0$ for the Pt/Fe bilayer. Since the non-zero value of $\Delta \xi_{\rm{FL}}^E$ arises from the $d_\mathrm{Pt}$-dependent component of the field-like torque, the opposite sign of $\Delta \xi_{\rm{FL}}^E$ shows that the sign of the field-like torque due to the bulk spin Hall effect in the Pt layer is opposite in the Pt/Ni and Pt/Fe bilayers. We also note that the field-like torque efficiency $\xi_{\rm{FL}}^E$ in Table~\ref{result} shows that the magnitude of the field-like torque in the Pt/Fe bilayer is clearly larger than that in the Pt/Ni bilayer. Since the above result is a rough estimation, we only focus on the sign and relative magnitude of the field-like torque below.

The interfacial spin-orbit coupling plays an important role in the generation of the field-like torque due to the bulk spin Hall effect, as well as that arising from the interface. 

In the model of the spin-orbit torques due to the bulk spin Hall effect, the real part of the spin-mixing conductance $\text{Re}[G_{\mathrm{Pt} / \mathrm{FM}}^{\uparrow \downarrow}]$ contributes to the damping-like torque, whereas the imaginary part $\text{Im}[G_{\mathrm{Pt} / \mathrm{FM}}^{\uparrow \downarrow}]$ contributes to the field-like torque~\cite{PhysRevB.87.174411}. The reason for this is that the field-like torque due to the spin Hall effect arises from the reflection of the spin current at the interface, while the damping-like torque arises from the injection of the spin current. Although $\text{Im}[G_{\mathrm{Pt} / \mathrm{FM}}^{\uparrow \downarrow}]$ has been believed to be negligible compared to $\text{Re}[G_{\mathrm{Pt} / \mathrm{FM}}^{\uparrow \downarrow}]$ in metallic films, recent studies have shown non-negligible $\text{Im}[G_{\mathrm{Pt} / \mathrm{FM}}^{\uparrow \downarrow}]$ in such systems~\cite{PhysRevB.89.174424,APL104.082407,PhysRevB.91.214416,PhysRevApplied.13.054011}. Since sizable $\text{Im}[G_{\mathrm{Pt} / \mathrm{FM}}^{\uparrow \downarrow}]$ appears when a spin current reflected at the interface experiences a large angle rotation of its spin direction due to interfacial spin-orbit coupling~\cite{PhysRevB.89.174424}, the stronger field-like torque in the Pt/Fe film, compared with that in the Pt/Ni film is consistent with the stronger spin-orbit coupling at the Pt/Fe interface.

The opposite sign of the field-like torque in the Pt/Ni and Pt/Fe bilayers indicates that the sign of $\text{Im}[G_{\mathrm{Pt} / \mathrm{FM}}^{\uparrow \downarrow}]$ is opposite in these devices. 

Here, the sign of the imaginary part of the spin mixing conductance can be different depending on the choice of the FM layer. For a model where all of the Fermi surfaces are spherical and the same size, the real and imaginary parts of the spin-mixing conductance due to spin-dependent reflection arising from a spin-dependent potential at the interface are expressed as~\cite{PhysRevB.87.174411,PhysRevB.91.144412} 
\begin{widetext}
\begin{equation}
\operatorname{Re}[G_{\mathrm{Pt} / \mathrm{FM}}^{\uparrow \downarrow}]= \frac{e^2 k_\mathrm{F}^2}{2\pi h}  \left( \frac{1}{2}+\frac{u^{\uparrow} u^{\downarrow}}{2\left(u^{\uparrow}+u^{\downarrow}\right)}\left[u^{\downarrow} \ln \left(\frac{u^{\downarrow^{2}}}{1+u^{\downarrow^{2}}}\right) +u^{\uparrow} \ln \left(\frac{u^{\uparrow^{2}}}{1+u^{\uparrow^{2}}}\right)\right]\right), \label{mixre}
\end{equation}
\begin{equation}
\operatorname{Im}[G_{\mathrm{Pt} / \mathrm{FM}}^{\uparrow \downarrow}]=\frac{e^2 k_\mathrm{F}^2}{2\pi h}  \left( \frac{u^{\uparrow} u^{\downarrow}}{2\left(u^{\uparrow}+u^{\downarrow}\right)}\left(u^{\downarrow}\left[\pi-2 \tan ^{-1} u^{\downarrow}\right]-u^{\uparrow}\left[\pi-2 \tan ^{-1} u^{\uparrow}\right]\right)\right).\label{mixim}
\end{equation}
\end{widetext}
Here, $u^{\uparrow}$ and $u^{\downarrow}$ represent the strength of the spin dependent potential at the interface, where $\uparrow$ and $\downarrow$ refer to majority and minority electrons respectively. Equations~(\ref{mixre}) and (\ref{mixim}) show that the sign of $\operatorname{Re}[G_{\mathrm{Pt} / \mathrm{FM}}^{\uparrow \downarrow}]$ is always positive. In contrast, the sign of $\operatorname{Im}[G_{\mathrm{Pt} / \mathrm{FM}}^{\uparrow \downarrow}]$ depends on  the relative strength between the spin dependent potentials $u^\uparrow$ and $u^\downarrow$. This indicates that the the sign of the field-like torque due to the spin Hall effect depends on the electronic structure of the FM layer.

\section{IV. conclusion}
In summary, we investigated the damping-like and field-like torque efficiencies in the Pt/Ni and Pt/Fe bilayers with different Pt-layer thicknesses to reveal the origin of the spin-orbit torques in the Pt-based structures.
We found that the damping-like torque efficiency $\xi_{\rm{DL}}^E$ is enhanced by increasing $d_\mathrm{Pt}$ in both Pt/Ni and Pt/Fe bilayers. The result also shows that the magnitude of $\xi_{\rm{DL}}^E$ is almost identical in the Pt/Ni and Pt/Fe bilayers despite the stronger spin memory loss at the Pt/Fe interface. These results suggest that although the $d_\mathrm{Pt}$ dependence of $\xi_{\rm{DL}}^E$ is consistent with the damping-like torque due to the bulk spin Hall effect in the Pt layer, the Pt/Fe interface also contributes to the damping-like torque in the Pt/Fe bilayer. The non-negligible contribution from the interface to the damping-like torque is consistent with that the spin-orbit coupling at the Pt/Fe interface is stronger than that at the Pt/Ni interface. Although a part of the damping-like torque in the Pt/Fe bilayer originates from the interface, the origin is unclear from the result; the intrinsic spin-orbit torque and/or the interfacial spin-orbit scattering may play an important role in the interface generation of the damping-like torque. 

In contrast to the damping-like torque, whose magnitude and sign are almost identical in the Pt/Ni and Pt/Fe bilayers, the field-like torque strongly depends on the choice of the FM layer. In particular, the sign of the $d_\mathrm{Pt}$-dependent component of the field-like torque efficiency $\xi_{\rm{FL}}^E$ is opposite between the Pt/Ni and Pt/Fe bilayers. This indicates that the direction of the field-like torque due to the bulk spin Hall effect in the Pt layer is opposite between the Pt/Ni and Pt/Fe bilayers, which is attributed to the opposite sign of the imaginary part of the spin-mixing conductance. The magnitude, as well as the sign, of the field-like torque is also clearly different in the Pt/Fe and Pt/Ni bilayers. The stronger field-like torque in the Pt/Fe bilayer indicates that the imaginary part of the spin-mixing conductance in the Pt/Fe bilayer is larger than that in the Pt/Ni bilayer, which is also consistent with the stronger interfacial spin-orbit coupling in the Pt/Fe bilayer.

Previous studies have demonstrated that the interfacial spin-orbit coupling is also strong in Pt/Co bilayers. However, the role of the bulk and interfacial spin-orbit coupling in the generation of the spin-orbit torques is clearly different between the Pt/Co and Pt/Fe bilayers. 
The dominant source of the damping-like torque in the Pt/Co bilayer is the bulk spin Hall effect in the Pt layer, which is the same as that in the Pt/Ni and Pt/Fe bilayers~\cite{PhysRevApplied.13.034038,PhysRevApplied.13.054014,PhysRevB.101.060405}. In contrast, the field-like torque in the Pt/Co bilayer is mainly generated by the Pt/Co interface; the contribution from the bulk spin Hall effect to the field-like torque is negligible in the Pt/Co bilayer. This is different from the field-like torque in the Pt/Fe bilayer, where the bulk spin Hall effect plays an important role, even though both films show the strong interfacial spin-orbit coupling. 
The different mechanisms responsible for the spin-orbit torques in the Pt/FM bilayers show that the electronic structure of the FM layer, as well as the interfacial spin-orbit coupling, plays an important role in the generation of the spin-orbit torques.

\section{ACKNOWLEDGMENTS}

\begin{acknowledgments}
This work was supported by JSPS KAKENHI Grant Numbers 19H00864, 19K22131, Canon Foundation, Asahi Glass Foundation, Kao Foundation for Arts and Sciences, JGC-S Scholarship Foundation, and Spintronics Research Network of Japan (Spin-RNJ). 
\end{acknowledgments}

\end{document}